\documentclass[twocolumn,showpacs,preprintnumbers,amsmath,amssymb,prl,aps]{revtex4-1}

\usepackage{graphicx}
\usepackage{bm}
\usepackage{float} 
\usepackage{multirow}
\usepackage{amsmath}
\usepackage{gensymb}
\usepackage{epsfig, epstopdf}
\usepackage[usenames, dvipsnames]{color}
\usepackage{xcolor, soul}
\begin{document}
\title{Anomalies in the temperature evolution of the Dirac states in a topological crystalline insulator SnTe}

\author{Ayanesh Maiti$^{1,2}$}
\author{Ram Prakash Pandeya$^1$}
\author{Bahadur Singh$^1$}
\author{Kartik K. Iyer$^1$}
\author{A. Thamizhavel$^1$}
\author{Kalobaran Maiti$^1$}

\affiliation{$^1$Department of Condensed Matter Physics and Materials Science, Tata Institute of Fundamental Research, Homi Bhabha Road, Colaba, Mumbai - 400005, India}
\affiliation{$^2$Department of Physics, Indian Institute of Science, Bangalore - 560012, India}


\begin{abstract}
Discovery of topologically protected surface states, believed to be immune to weak disorder and thermal effects, opened up a new avenue to reveal exotic fundamental science and advanced technology. While time-reversal symmetry plays the key role in most such materials, the bulk crystalline symmetries such as mirror symmetry preserve the topological properties of topological crystalline insulators (TCIs). It is apparent that any structural change may alter the topological properties of TCIs. To investigate this relatively unexplored landscape, we study the temperature evolution of the Dirac fermion states in an archetypical mirror-symmetry protected TCI, SnTe employing high-resolution angle-resolved photoemission spectroscopy and density functional theory studies. Experimental results reveal a perplexing scenario; the bulk bands observed at 22 K move nearer to the Fermi level at 60 K and again shift back to higher binding energies at 120 K. The slope of the surface Dirac bands at 22 K becomes smaller at 60 K and changes back to a larger value at 120 K. Our results from the first-principles calculations suggest that these anomalies can be attributed to the evolution of the hybridization physics with complex structural changes induced by temperature. In addition, we discover drastically reduced intensity of the Dirac states at the Fermi level at high temperatures may be due to complex evolution of anharmonicity, strain, etc. These results address robustness of the topologically protected surface states due to thermal effects and emphasize importance of covalency and anharmonicity in the topological properties of such emerging quantum materials.
\end{abstract}

\maketitle


Topological insulators\cite{TI} are bulk insulators with symmetry protected metallic surface states which form Dirac cone energy dispersions within the bulk energy bandgap. The topological protection makes these Dirac Fermionic states robust against weak disorder, temperature, various surface effects, etc. It is proposed that such properties can be exploited to realize exotic physics involving Majorana fermions \cite{majorana}, magnetic monopoles \cite{monopole} as well as application in advanced electronic, spintronics, and quantum technologies. In these materials, the topological protection is realized by time-reversal symmetry in strongly spin-orbit coupled systems. Various experiments reported instability in the properties of these surface states with aging and/or deposition/adsorption of foreign elements on the surface \cite{aging-Hasan,aging-Benia,Deep-SREP,aging-Bianchi,PbSnTe-InGap-STM}. It is shown that covalent bond of the surface atoms with the adsorbed elements significantly affects the topological order \cite{aging-Kong,Deep-JESRP}. While impurity induced effects are being studied well, the role of thermodynamic parameters on the stability of Dirac states remained relatively unexplored despite its importance for material engineering.

\begin{figure}
\vspace{-6ex}
 \begin{center}
 \includegraphics[width=0.5\textwidth]{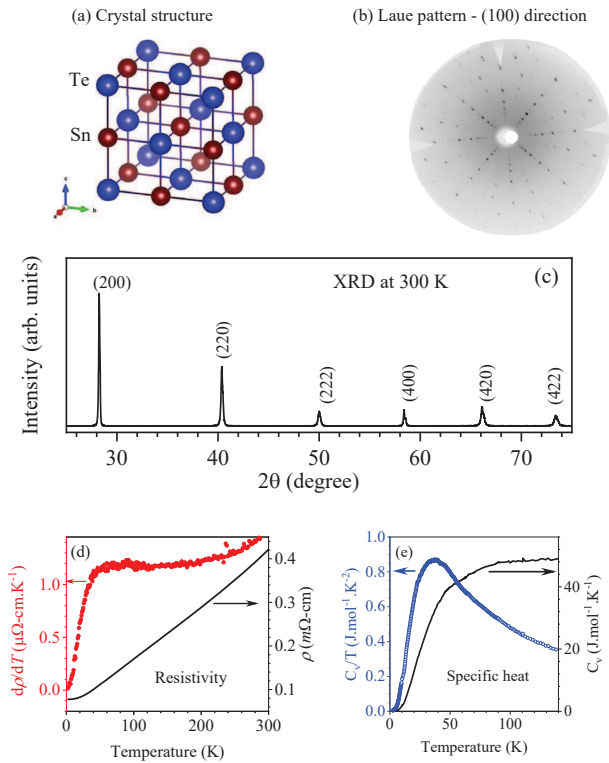}
 \end{center}
\vspace{-2ex}
\caption{(a) Crystal structure of SnTe. (b) Laue diffraction pattern, taken on the (100) surface. (c) Powder $x$-ray diffraction pattern at 300 K. (d) Resistivity, $\rho$ (line) and $d\rho/dT$ (symbols). (e) Specific heat, $C_v$ (line) and $C_v/T$ (symbols).}
\label{Fig1Str}
\end{figure}

In topological crystalline insulators (TCIs), crystalline symmetries such as mirror symmetry protects the surface states. Thus, TCIs are inherently expected to be more sensitive to the thermodynamic parameters such as temperature, pressure, etc. Angle-resolved photoemission spectroscopy (ARPES) studies established topological crystalline order in tin telluride, SnTe \cite{SnTe-ARPES-nphys,PbSnTe-ARPES-ncomm,SnTe-ARPES111-PRB}. Pb$_{1-x}$Sn$_x$Se also show TCI phase for $0.2\leq x \leq 0.45$, where heating leads to topologically trivial state above 250 K \cite{PbSnSe}. SnTe is different having no chemical substitutional disorder like PbSnSe-system. It forms in rock salt crystal structure at room-temperature (Fig. \ref{Fig1Str}(a)) and is a good candidate for thermoelectric applications \cite{Kanishka-SnTe}. Various studies \cite{Rhomb-JPSJ75,Arindam-AIP,Rhomb-PRL76,Phonon-PRL2018} reported rhombohedral distortion at low temperatures. The transition temperature increases with the increase in carrier density arising from Sn-deficiency and/or defects in the system. Evidently, SnTe lies in the proximity of complex structural phase boundary and provides an ideal platform to test the robustness of the topologically ordered states on varied conditions. We have grown high quality single crystals of SnTe and studied the electronic structure employing ARPES and density functional theory (DFT)-based calculations. The experimental results reveal puzzling spectral changes with temperature.


High quality single crystalline samples were grown by modified Bridgman method and characterized through measurement of varied structural and electronic properties  \cite{Ayanesh-bulk}. The Laue diffraction pattern shown in Fig. \ref{Fig1Str}(b) exhibits bright sharp spots implying very high quality crystalline order. A powder $x$-ray diffraction pattern is shown in Fig. \ref{Fig1Str}(c). Sharp distinct peaks related to the rock salt structure are observed suggesting single phase with no signature of impurity. In Fig. \ref{Fig1Str}(d), we show the resistivity ($\rho$) exhibiting metallic behavior in the whole temperature range studied. $d\rho/dT$ is nearly temperature independent down to about 50 K and then decreases sharply due to strong electron-phonon coupling \cite{Rhomb-PRL76,Phonon-PRL2018}. Signature of structural transition (a sharp peak) is not observed in the $d\rho/dT$) plot. In Fig. \ref{Fig1Str}(e), we show the specific heat, $C_v$ and $C_v/T$ with temperature exhibiting typical behavior due the phonon-contributions. $C_v$ is negligibly small at low temperatures indicating weak contribution from the conduction electrons. The specific heat data do not show  signature of phase transition. All these results suggest good crystallinity and high quality of the sample having absence of structural transition down to the lowest temperature studied.

ARPES measurements were carried out using a state-of-the-art DA30L electron analyser and monochromatic UV source, VUV5k ($h\nu$ = 21.2 eV) from Scienta Omicron at a pressure of about 7$\times$10$^{-11}$ torr. The energy resolution was set to 5 meV. Samples were cleaved in the spectrometer chamber just before the measurements and the sample temperature was varied using a closed cycle helium cryostat from Advanced Research systems, USA. Use of DA30L analyser allowed us to do all the measurements at the same experimental geometry after the initial sample alignment after cleaving.

Electronic structure calculations were carried out within the DFT framework \cite{dft} with projector augmented wave (PAW) potentials as implemented in the Vienna {\it ab-initio} simulation package (VASP) \cite{paw,vasp}. The generalized gradient approximation (GGA) with Perdew-Burke-Ernzerhof parameterization was used to consider exchange-correlation effects \cite{pbe}. We used an energy cut-off of 208 eV for the plane-wave basis set and a $\Gamma$-centered $13 \times 13 \times 13$ $k$ mesh for the bulk Brillouin zone sampling.  A tolerance of $10^{-8}$ eV was used for electronic energy minimization. A tight-binding model with atom-centered Wannier functions was generated using the VASP2WANNIER90 interface\cite{w90}, while the topological surface states were obtained within the iterative Green’s function method using the Wanniertools package\cite{wtools}.


\begin{figure}
\centering
\includegraphics[width=0.5\textwidth]{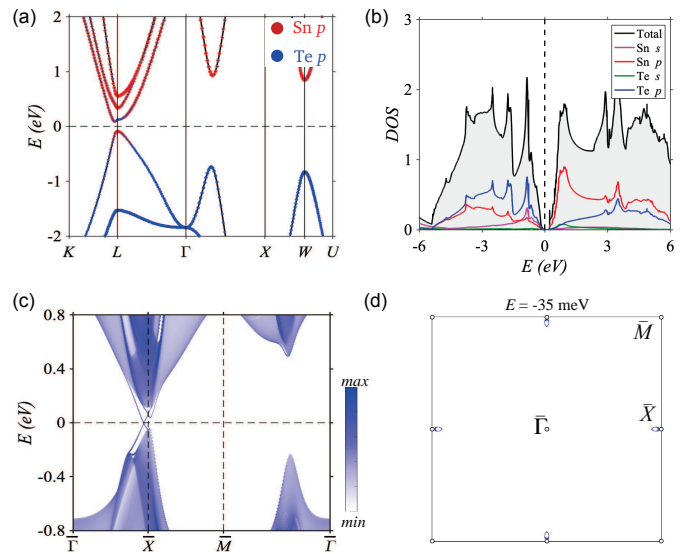}
\caption{(a) Orbital-resolved band structure of SnTe. Sn 5$p$ and Te 5$p$ states compositions are shown in red and blue, respectively. (b) Sn 5$s$ (pink), Sn 5$p$ (red), Te 5$s$ (green) and Te 5$p$ (blue) PDOS. (c) Band structure of (001) surface of SnTe showing a Dirac cone close to  $\overline{X}$ point. (d) Constant energy contours at 35 meV below the Fermi level.}
\label{Fig2DFT}
\end{figure}

\begin{figure}
 \begin{center}
 \includegraphics[width=0.5\textwidth]{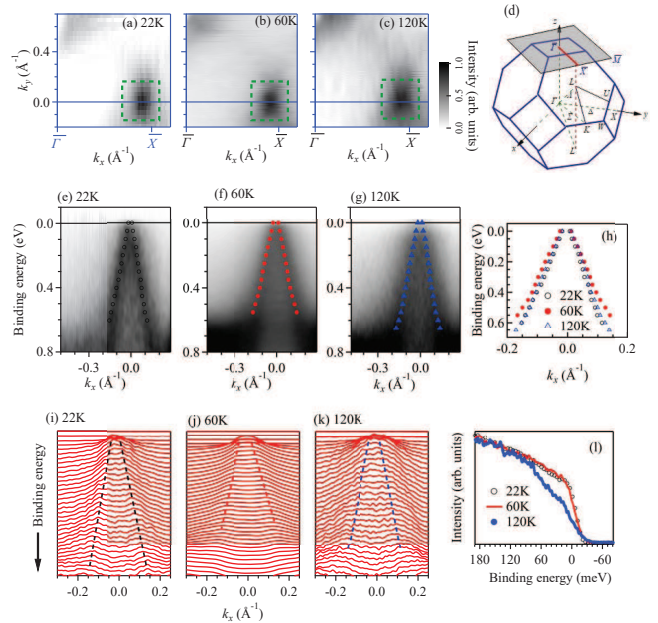}
 \end{center}
\vspace{-28ex}
 \caption{Fermi surface at (a) 22 K, (b) 60 K, and (c) 120 K. (d) A sketch of the reciprocal lattice and the surface Brillouin zone. The ARPES data obtained from a horizontal cut across the Fermi pocket (blue line) at (e) 22 K, (f) 60 K, and (g) 120 K. The corresponding momentum distribution curves (MDCs) at (i) 22 K, (j) 60 K and (k) 120 K. Dashed lines are the guide to eye. The symbols are the peak positions derived by fitting the MDCs. (h) Traces of the energy bands (symbols) at different temperatures obtained by fitting the MDCs. (l) Energy distribution curves (EDCs) obtained by momentum integration within the green box shown in (a), (b), and (c).}
 \label{Fig3ARPES}
\end{figure}

The calculated bulk band structure of SnTe is shown in Fig. \ref{Fig2DFT}(a). The valence and conduction bands are well separated in energy at all $k$ points without crossing the Fermi level $\epsilon_F$, showing an insulating ground state. A direct bandgap of 214 meV is observed at $L$ point in the bulk Brillouin zone; a sketch of the bulk Brillouin zone and its projection to (001) surface is shown in Fig. \ref{Fig3ARPES}(d). The orbital character of the bands exhibits band inversion at $L$ point which is a signature of topological order consistent with earlier studies \cite{SnTeDFT-ncomm}.

We present the calculated partial density of states (PDOS) corresponding to Sn 5$s$, Sn 5$p$, Te 5$s$ and Te 5$p$ orbitals in Fig. \ref{Fig2DFT}(b). All the contributions are spread over the entire energy range shown in Fig. \ref{Fig2DFT}(b), indicating a mixed character of the bands. The spectral intensity just below $\epsilon_F$ is primarily contributed by the Te 5$p$ states. Sn 5$s$ PDOS exhibit similar energy distribution in this regime and more intense than Sn 5$p$ PDOS due to strong Sn 5$s$-Te 5$p$ hybridization. The states above $\epsilon_F$ are essentially Sn 5$p$ PDOS with small contributions from other orbitals. The feature near $\epsilon_F$ in the conduction band has similar energy distributions for Sn 5$p$ and Te 5$s$ states reflecting strong Sn 5$p$-Te 5$s$ hybridization. Evidently, $sp$ hybridization plays a key role in the band inversion \cite{SnTe-DFT-npj} while the $pp$ mixing is prominent in the features away from $\epsilon_F$.

Figure \ref{Fig2DFT}(c) shows the (001) surface band structure of SnTe. A Dirac cone state is seen along $\overline{\Gamma} - \overline{X}$ mirror direction in the vicinity of $\overline{X}$ point, establishing the topological behavior of the system. Notably, the projection of the bulk $L$ points where the system has an inverted band ordering lies on $\overline{\Gamma}-\overline{X}$ vector of the surface Brillouin zone. Owing to the nontrivial mirror Chern number, an even number of topological states should exist on the $\overline{\Gamma} - \overline{X}$ line as seen in our results; the location of the Dirac cone is denoted by $\Lambda$ \cite{SnTe-ARPES-nphys}. We find four Dirac cones located on the four symmetry equivalent $\overline{\Gamma} - \overline{X}$ lines in the (001) surface Brillouin zone consistent with earlier studies. This is further resolved in the constant energy contours at 35 meV below $\epsilon_F$ in Fig. \ref{Fig2DFT}(d) where four energy pockets are seen on near $\overline{X}$ around the Dirac point.

In Fig. \ref{Fig3ARPES}, we show the ARPES results collected at different temperatures. The use of the DA30L analyzer helped to map the Fermi surface without changing the sample position and/or experimental geometry. The Fermi surface at 22 K is shown in Fig. \ref{Fig3ARPES}(a). It appears at $\Lambda$-point consistent with earlier reports \cite{SnTe-ARPES-nphys,PbSnTe-ARPES-ncomm,SnTe-ARPES111-PRB} as well as our theoretical results discussed above. The shape of the Fermi surface and it's location in the surface Brillouin zone remain unchanged with heating upto 120 K as demonstrated in Figs. \ref{Fig3ARPES}(b) and (c) where the Fermi surfaces at 60 K and 120 K are plotted. These results are in line with the expected behavior as no structural transition is found in our sample and the Dirac Fermions are protected from small thermal effects.

The energy bands obtained from the cut along $\overline{\Gamma}$-$\overline{X}$ line are shown in Fig. \ref{Fig3ARPES}(e), (f) and (g) for the temperatures 22 K, 60 K and 120 K, respectively. The corresponding momentum distribution curves (MDCs) are shown in Fig. \ref{Fig3ARPES}(i), (j) and (k). Distinct Dirac bands are observed at all the temperatures. The energy position of the bulk bands and the slope of Dirac bands exhibit anomalous change with different temperatures. To study the slope change with temperature, we have extracted peak positions by fitting MDCs. The results are shown by symbols in Figs. \ref{Fig3ARPES}(e), (f) and (g) superimposed over the image plots exhibiting good description of experimental data. The data at all the temperatures are superimposed over each other in Fig. \ref{Fig3ARPES}(h) for comparison. The Dirac point in all the cases appears just above $\epsilon_F$ consistent with earlier observation \cite{PbSnTe-ARPES-ncomm}. The slope of the Dirac bands at 60 K is clearly smaller than the slopes at other temperatures.

The intensity of the Dirac bands near $\epsilon_F$ also shows unusual behavior. To demonstrate this, we superimposed the momentum ($k$-points within the square box)-integrated spectral intensities in Fig. \ref{Fig3ARPES}(l). The spectral functions near $\epsilon_F$ look very similar at 22 K and 60 K apart from the changes due to the Fermi-Dirac distribution function. The 120 K data exhibit a large depletion of intensity near $\epsilon_F$; a sharp deviation from the expected behavior of topologically protected states immune to such small thermal perturbations.

\begin{figure}
 \centering
 \includegraphics[width=0.45\textwidth]{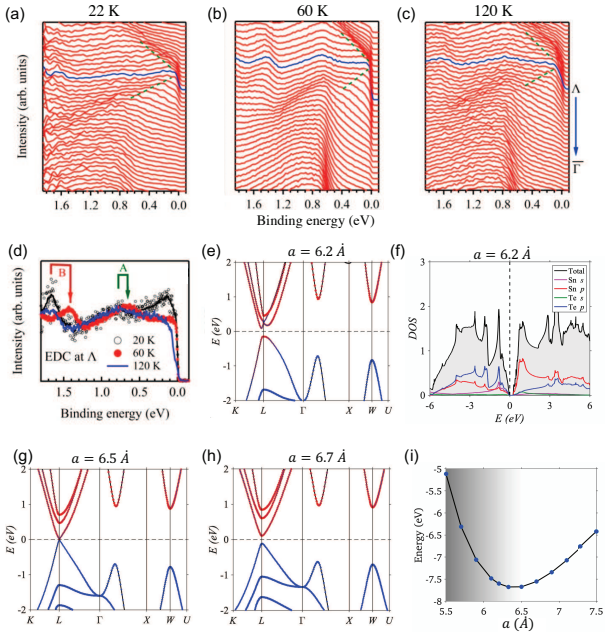}
\caption{Energy distribution curves (EDCs) at (a) 22 K, (b) 60 K, and (c) 120 K. Dashed lines are guide to eye. (d) EDCs at $\Lambda$-point at 22 K (open circles), 60 K (closed circles) and 120 K (line). Solid black line is a smooth curve of the 22 K data. (e) Band structure calculated using the lattice constant $a$ = 6.2 \AA\ and (f) the corresponding Sn 5$s$, Sn 5$p$, Te 5$s$ and Te 5$p$ PDOS. Band structure for (g) $a$ = 6.5 \AA\ and (h) $a$ = 6.7 \AA. (i) Total energy as a function of $a$(\AA). The shaded region shows the inverted band structure region.}
 \label{Fig4EDC}
\end{figure}

To probe this perplexing scenario further, we investigate the energy distribution curves (EDCs) along the $k$-vector, $\Gamma \Lambda X$ in Fig. \ref{Fig4EDC}. The bands close to $\epsilon_F$ are the surface Dirac bands shown by dashed lines in the figure. Two anomalies are evident in the spectra. (i) The bulk bands at 22 K (bands between 0.7 - 0.8 eV and at higher binding energies) shifted towards $\epsilon_F$ at 60 K and then moved back to higher binding energies at 120 K. (ii) The 22 K spectra are dominated by the intensities of the Dirac bands. The scenario got reversed at other temperatures exhibiting Dirac bands much weaker than the bulk bands. For better demonstration, we superimposed the EDCs at $\Lambda$ at 22 K, 60 K and 120 K in Fig. \ref{Fig4EDC}(d). The similarity of the bulk peak energies at 22 K and 120 K and a shift towards $\epsilon_F$ at an intermediate temperature of 60 K is evident as marked by A and B. The intensity of the Dirac Fermionic peak reduces significantly at 60 K relative to 22 K data. The experimental results at 120 K exhibit a puzzling sharp drop in intensity at $\epsilon_F$.

The thermal compression of the system due to cooling will enhance the hybridization. This is expected to affect the bulk bandwidth and the band inversion significantly. In order to investigate this scenario, we calculated the bulk band structure as a function of lattice parameter $a$. The results for $a$ = 6.2 \AA, 6.5 \AA\ and 6.7 \AA\ are shown in Figs. \ref{Fig4EDC}(e), (g) and (h), respectively. While the overall band structure looks quite similar to the results obtained for the room temperature lattice parameters (see Fig. \ref{Fig2DFT}), the description of the bands at $L$-point is significantly different. In Fig. \ref{Fig4EDC}(e), the dispersion of the valence band in the vicinity of $L$ becomes shallower than the data in Fig. \ref{Fig2DFT}(a) and the conduction bands show crossing of the bands at $L$. This implies a larger overlap of the valence and conduction bands at smaller $a$ and stronger band inversion effect. The character of the bands shown in the PDOS plot in Fig. \ref{Fig4EDC}(f) indicate a strong mixing as also observed in Fig. \ref{Fig2DFT}(b). The band structure results for $a$ = 6.5 \AA\ and 6.7 \AA\ show a cross over from band inversion regime to an atomic limit band ordering at $a \sim$ 6.5 \AA. The total energy calculation shown in Fig. \ref{Fig4EDC}(i) exhibit minima for the $a$-range 6.2 - 6.7 \AA. All these results demonstrate a stronger band inversion if the sample is cooled from room temperature. In addition, we observe signature of indirect bandgap at $a$ = 6.2 \AA. Thus, the optical properties at this limit needs involvement of lattice degrees of freedom.

The theoretical results discussed above suggests that the thermal compression enhances flatness of the top of the valence band and the bottom of conduction band moves away from the $L$-point. Therefore, the slope of the Dirac bands connecting the valence and conduction bands will become smaller at lower temperatures. Such change in slope due to the thermal compression has been verified in our calculations. This is also reflected in Fig. 4; the valence band edge below 1.5 eV moves significantly towards $\epsilon_F$ at 60 K relative to the 120 K data. While these results are consistent with the experimental data at 120 K and 60 K, the 22 K data show a reversed scenario. To gain an insight of this scenario, let us note that significant Sn-deficiency in SnTe leads to a structural transition from the cubic (Space group $Fm\overline{3}m$) to a rhombohedral structure (Space group $R3m$) at low temperatures \cite{Rhomb-JPSJ75,Arindam-AIP,Rhomb-PRL76, SnTe-QuOscARPES-distortion-PRB20}. The transition temperature reduces with the reduction in career concentration that suggests that SnTe is in the proximity of structural transition phase boundary which often has implications in the electronic properties hosting hidden order as found in other materials \cite{Ca122-ARPES}. A deviation from the cubic structure will reduce the hopping integral, $t_{ij} = \langle j|H|i\rangle$ and hence the width of valence and conduction bands. Therefore, the energy eigenvalues near the top of the valence band will shift towards higher binding energies which might be a reason for the shift of the bulk peaks at 22 K. In such a case, the bulk band gap will increase leading to an enhancement of the slope of the Dirac bands.

While these results provide a reasonable description of the anomalies observed in the band structure, the sharp reduction in spectral intensity at $\epsilon_F$ with the increase in temperature is still a puzzle. The photoemission cross-section, $\sigma(\epsilon)$ at 21.2 eV photon energy is 0.065 (Sn 5$s$), 1.173 (Sn 5$p$), 0.022 (Te 5$s$) and 4.758 (Te 5$p$) \cite{yeh-lindau}. Thus, a change in hybridization will change the cross-section of the photo-excitation of the eigenstates. However, such a description cannot capture the experimental scenario exhibiting anomalous changes close to $\epsilon_F$.

It is to note here that the phase transition in SnTe is of displacive-type \cite{pawley-PRL66,CONeill-PRB17} which is reflected in the transport properties enabling to track the structural transition by monitoring the peak in $d\rho/dT$ \cite{Rhomb-PRL76}. The transverse optical modes soften at the transition temperature and the phonon instability appears as negative phonon frequencies \cite{Phonon-PRL2018,CONeill-PRB17}. These properties have been exploited to engineer enhanced thermoelectric properties of this material \cite{Kanishka-SnTe}. The Debye temperature extracted from the analysis of our specific heat data is close to 100 K. A change in anharmonicity and/or many-body dipole-dipole screening plays important role in this material \cite{CONeill-PRB17}. Effect due to strain \cite{Vidya-NComm} and coexistence of massless Dirac fermions with massive ones have been observed \cite{Vidya-Science}. These observations suggest that anharmonicity in the phonon modes have significant influence in the symmetry protection of the surface states close to $\epsilon_F$ while the states at higher energies will be relatively less affected. Evidently, these results open up an unknown paradigm for extensive future studies which has immense implication in realizing exotic fundamental science and advanced technology at operable temperatures.


In summary, we have studied the temperature evolution of the Dirac cone states in an archetypical topological crystalline insulator, SnTe employing high-resolution ARPES and density functional theory. The experimental results reveal anomalies in the energy shift of bulk bands and change of the slope of Dirac bands with the change in temperature. The analysis of the experimental data in conjunction with the theoretical first principle calculations suggest complex evolution of the hybridization and band inversion properties as a reason for the anomalous spectral evolution. In addition, we discover drastic reduction of the intensity of the Dirac states near the Fermi level with the increase in temperature. This is attributed to the complex evolution of the anharmonicity and strain in this system that influences the crystalline symmetry protection. These results provide an example of the importance of covalency and phonon anomalies in the topological properties of such emerging quantum materials.


Authors acknowledge support from the Dept. of Atomic Energy, Govt. of India under the project no. 12-R\&D-TFR-5.10-0100. AM acknowledges financial assistance from the Dept. of Science and Technology under the Kishore Vaigyanik Protsahan Yojana (KVPY) program and KM acknowledges support from BRNS, DAE, Govt. of India under the DAE-SRC-OI Award (grant no. 21/08/2015-BRNS/10977).

\end{document}